\documentclass[12pt]{article}
\usepackage{amsmath, amssymb, slashed, epsf, color, graphicx}
\usepackage{epstopdf}

\IfFileExists{srcltx.sty}{\usepackage[active]{srcltx}}

\newcommand{\vev}[1]{ \left\langle {#1} \right\rangle }
\newcommand{\bear}{\begin{array}}
\newcommand {\eear}{\end{array}}
\newcommand{\bea}{\begin{eqnarray}}
\newcommand{\eea}{\end{eqnarray}}
\newcommand{\beq}{\begin{equation}}
\newcommand{\eeq}{\end{equation}}
\newcommand{\bef}{\begin{figure}}
\newcommand {\eef}{\end{figure}}
\newcommand{\bec}{\begin{center}}
\newcommand {\eec}{\end{center}}
\newcommand{\non}{\nonumber}

\newcommand{\la}{\left\langle}
\newcommand{\ra}{\right\rangle}

\def\lrfp#1#2#3{ \left(\frac{#1}{#2} \right)^{#3}}
\def\GEV#1{10^{#1}{\rm\,GeV}}

\def\lrfp#1#2#3{ \left(\frac{#1}{#2} \right)^{#3}}

\begin{document}

\begin{titlepage}
\begin{center}

\hfill IPMU-13-0069 \\
\hfill \today

\vspace{1.5cm}
{\large\bf Neutrinos at IceCube from Heavy Decaying Dark Matter}

\vspace{2.0cm}
{\bf Brian Feldstein}$^{(a)}$,
{\bf Alexander Kusenko}$^{(a, b)}$, \\
{\bf Shigeki Matsumoto}$^{(a)}$,
and
{\bf Tsutomu T. Yanagida}$^{(a)}$

\vspace{1.5cm}
{\it
$^{(a)}${Kavli Institute for the Physics and Mathematics of the Universe (WPI), \\
University of Tokyo, Kashiwa, Chiba 277-8568, Japan} \\
$^{(b)}${ Department of Physics and Astronomy, \\
University of California, Los Angeles, CA 90095-1547, USA}
}

\vspace{1.5cm}
\abstract{
A monochromatic line in the cosmic neutrino spectrum would be a smoking gun signature of dark matter.
It is intriguing that the IceCube experiment has recently reported two
PeV neutrino events with energies that may be equal up to experimental
uncertainties, and which have a probability of being 
a background fluctuation estimated to be less than a percent. Here we explore prospects for these events to be the first indication of a monochromatic line signal from dark matter. While measurable annihilation signatures would seem to be impossible at such energies, we discuss the dark matter quantum numbers, effective operators, and lifetimes which could lead to an appropriate signal from dark matter decays. We will show that the set of possible decay operators is rather constrained, and will focus on several viable candidates which could explain the IceCube events; R-parity violating gravitinos, hidden sector gauge bosons, and singlet fermions in an extra dimension. In essentially all cases we find that a PeV neutrino line signal from dark matter would be accompanied by a potentially observable continuum spectrum of neutrinos rising towards lower energies.
}

\end{center}
\end{titlepage}
\setcounter{footnote}{0}

\section{Introduction}

The IceCube collaboration has very recently reported a detection of two neutrino events with energies of 1.1~PeV and 1.3~PeV
 in an energy range where no more than 0.01 background events was expected from atmospheric neutrinos~\cite{icecube0, icecube1, icecube2}. These are stated to be either electron neutrino charged current events, or neutral current events of any neutrino flavor. It is interesting that the two detected neutrinos have such similar energies, and indeed, most astrophysical sources are expected to produce power-law spectra-- in particular one might have expected to see additional 
events at 
around 6.8~PeV, where the detector sensitivity is enhanced by the Glashow resonance~\cite{Bhattacharya:2011qu}. The data may thus suggest a peak, or falloff, in the neutrino spectrum around 1~PeV. It is possible that such a spectrum could be produced by some astrophysical sources~\cite{Cholis:2012kq}--\cite{Liu:2012pf}, including intergalactic interactions of cosmic rays produced by blazars~\cite{Essey:2009zg}--\cite{Kalashev:2013vba}, but these models rely on some assumptions about the properties and evolution of the sources, as well as the intergalactic magnetic fields.

The IceCube observations raise a question of whether dark matter could be composed of relic particles whose decays or annihilations into neutrinos produce a feature in the neutrino spectrum at $\sim$PeV energy.  In this paper we will
concentrate on the possibility that this feature could actually be a monochromatic neutrino line. 
Similar to a line in the gamma ray spectrum, a line in the neutrino spectrum could be considered a ``smoking gun" signature for dark matter. Such line-like neutrino signatures from dark matter have been considered before \cite{Allahverdi:2009se}--\cite{Lindner:2010rr}, but in this paper we consider the possibility of obtaining such a signal at the PeV scale, where the dark matter particle cannot be a simple thermal relic. As we will show, the possibilities for obtaining a neutrino line signal from dark matter at such energies are highly constrained, but there are nevertheless various viable scenarios.  We should note that due to the low statistics in the present data,  power law spectra
from cascade annihilations or decays of dark matter into neutrinos might also give reasonable fits.  We limit ourselves here to the possibility
of a line signature since this is the most exciting case--  with further data, a line signature would directly point towards
a dark matter explanation, whereas a power law signature might be difficult to disentangle from astrophysical sources.

For dark matter with an annihilation cross section into monochromatic neutrinos saturating the unitarity limit, $\sigma_{\rm Ann} \leq 4 \pi/(m_{\rm DM}^2 v^2)$, the event rate expected at a neutrino telescope of fiducial volume $V$ and nucleon number density $n_{\rm N}$ is of order
\begin{eqnarray}
\Gamma_{\rm Events}
\sim
V \, L_{\rm MW} \, n_{\rm N} \, \sigma_{\rm N} \,
\left(\frac{\rho_{\rm DM}}{m_{\rm DM}}\right)^2
\langle \sigma_{\rm Ann} v \rangle
\lesssim 
1 {\rm\ per\ few\ hundred\ years},
\end{eqnarray}
where we have taken the neutrino-nucleon scattering cross section to be $\sigma_{\rm N} \sim 9 \times 10^{-34} \, {\rm cm^2}$
at $E_\nu \simeq 1.2$~PeV \cite{Gandhi:1998ri}, and the nucleon number density to be that of ice, $n_{\rm N} \simeq n_{\rm Ice} \simeq 5 \times 10^{23}$/cm$^3$. $\rho_{DM}$, $v$, and $L_{\rm MW}$ are the milky way dark matter density (taken near the Earth for the purpose of our estimate), the typical dark matter particle velocity, and the rough linear dimension of our galaxy, where these are fixed to be 0.4 GeV/cm$^3$, $10^{-3}$, and 10 kpc, respectively. The fiducial volume $V$ is set to be 1 km$^3$, which is roughly the size of the IceCube detector. We see that obtaining a neutrino line signal from dark matter annihilations at the PeV scale is essentially not possible.  In what follows we will therefore restrict ourselves to the possibility of a signal from dark matter decays.

For dark matter decays also, obtaining a neutrino line signal at the energies of interest here turns out to be challenging.
Indeed, suppose one wishes to mediate an appropriate decay via a simple dimension 4 operator such as $\mathcal{L} \supset \lambda \bar{\psi} L H$, where $\lambda$ is a coupling constant, $\psi$ is the dark matter particle, $L$ is a lepton doublet, and $H$ is the Higgs doublet. Then the decay rate to neutrinos is $\Gamma_{\rm DM} = \frac{\lambda^2}{16 \pi} m_{\rm DM}$. Similarly to the annihilation case above, we may estimate the event rate at a neutrino detector for $m_{DM} \simeq 1.2$~PeV as
\begin{eqnarray}
\Gamma_{\rm Events}
\sim
V \, L_{\rm MW} \, n_{\rm N} \, \sigma_{\rm N} \,
\frac{\rho_{\rm DM}}{m_{\rm DM}} \, \Gamma_{\rm DM}
\sim
\left(\frac{\lambda}{ 10^{-29}}\right)^2 \,
/{\rm\ year}.
\end{eqnarray}
Clearly an exceptionally tiny coupling is required to obtain an appropriate signal, and a certain amount of model building would appear necessary.

We may also consider whether or not higher dimension operators, suppressed by some large mass scale, could give more naturally small event rates. For higher dimension operators, however, it is a nontrivial constraint that in order to obtain a line signal, the decay final state must be two-body. Indeed, for many interesting operators, neutrinos appear in the gauge singlet combination $L H$, and although naively this could lead to a neutrino decay with $H$ replaced by its vacuum expectation value $v$, this tends not to be the dominant process due to the large dark matter masses under consideration. For example, if one considers the operator $\mathcal{L} \supset \phi (L H)^2/\Lambda$ for a scalar dark matter particle $\phi$, and with $\Lambda$ a heavy mass scale, then the square of the amplitude for a four-body decay with two neutrinos and two Higgses is larger than that for a two-body neutrino decay by a factor of $\sim (m_{\rm DM}/v)^4$. For heavy dark matter masses, phase space suppressions for multi-body final states are not enough to prevent the four-body decay from being by far dominant.

In this paper we will comprehensively discuss effective operators which could mediate the decays of heavy dark matter particles into monochromatic neutrino lines, and we will find that only a handful of operators are viable. Several of these stand out as being particularly interesting, and we will discuss possible models for them in detail. These will include the cases of gravitino dark matter, with a mass motivated by the recent $125$ GeV Higgs discovery, a hidden gauge boson with an extremely small mixing with hypercharge, and a singlet fermion in an extra dimension. In each case we will discuss simple ways in which an appropriately long lifetime for the dark matter particle may be obtained in order to explain the IceCube data. In the gravitino case, the decay operator may be naturally suppressed by the scales of R-parity violation and lepton number violation. In the gauge boson case, the kinetic mixing with hypercharge may be suppressed by the scale of non-abelian gauge symmetry breaking in the hidden sector, as well as the breaking of grand unified symmetry in the visible sector. In the extra dimensional model, the required highly suppressed coupling may be produced naturally by an exponentially small wave-function factor.

The outline of this paper is as follows: In section \ref{sec: icecube} we will review the nature of the PeV IceCube neutrino events, as well as discuss the lifetime and mass of dark matter particles which may be able to explain them. In section \ref{sec: effective operators} we will discuss in general the effective operators which might be able to lead to an appropriate dark matter decay. Models yielding some of these operators will be discussed in section \ref{sec: models}. An interesting conclusion of our analysis will be that in essentially all cases, a monochromatic neutrino line would be accompanied by an appreciable
continuum spectrum of neutrinos rising towards lower energies. The prospects for detecting such a signature will be discussed in section \ref{sec: future}.

\section{IceCube Events}
\label{sec: icecube}

Before going on to discuss effective operators which could mediate the decays of heavy dark matter particles into monochromatic neutrino lines, we summarize the situation with the PeV neutrino events which have recently been reported by the IceCube collaboration. According to a plot in reference~\cite{icecube1}, the exposures at the energies of the two events turn out to be $4.4 \times 10^9$ [m$^2$ s sr] and $5.9 \times 10^9$ [m$^2$ s sr] for the 1.1 and 1.3
 PeV events, respectively, assuming that both the two events were caused by electron neutrinos.\footnote{It is possible that one or both events could have been caused by neutral current interactions of arbitrary flavor, but in such cases
one would expect the event energies to be much more spread out.} 
It follows that the total flux may be estimated to be
\begin{eqnarray}
{\cal F} \simeq 4.0 \times 10^{-14} \
[{\rm cm}^{-2} \ {\rm s}^{-1} \ {\rm sr}^{-1}].
\label{eq: total flux exp}
\end{eqnarray}

The observed neutrino event energies imply a mass for the dark matter particle of about 2.4~PeV, while the neutrino flux can be related to the lifetime for dark matter neutrino decays. When the mass of the decaying dark matter particle is assumed to be 2.4~PeV, the predicted flux of line neutrinos is estimated to be
\begin{eqnarray}
E_\nu^2 \frac{d{\cal F}}{dE_\nu} \simeq
9.5 \times 10^{-3} N_\nu
\left( \frac{10^{29} \ {\rm s}}{\tau_{\rm DM}} \right)
\delta(E_\nu - m_{\rm DM}/2) \
[{\rm GeV} \ {\rm cm}^{-2} \ {\rm s}^{-1} \ {\rm sr}^{-1}],
\end{eqnarray}
where $E_\nu$, $m_{\rm DM}$, $\tau_{\rm DM}$, and $N_\nu$ are the neutrino energy, the dark matter mass, its lifetime, and the number of neutrinos produced in each decay, respectively. Here the NFW profile was used for the dark matter density in our galaxy, and we have adopted profile parameters with a critical radius of $r_c = 20$ kpc, and a density at the solar-system of $\rho_\odot =$ 0.39~GeV/cm$^3$~\cite{Esmaili:2012us}. The total flux is then given by
\begin{eqnarray}
{\cal F} \simeq
0.76 N_\nu \times 10^{-14}
\left( \frac{10^{29} \ {\rm s}}{\tau_{\rm DM}} \right) \
[{\rm cm}^{-2} \ {\rm s}^{-1} \ {\rm sr}^{-1}].
\label{eq: total flux theory}
\end{eqnarray}

By comparing this prediction with the flux in equation~(\ref{eq: total flux exp}), we find that the  the lifetime of the dark matter particle must have the following value in order to explain the data:
\begin{eqnarray}
\tau_{\rm DM} &\simeq& 1.9 N_\nu \times 10^{28} \ {\rm s}.
\label{tau}
\end{eqnarray}
We have thus found that a decaying dark matter particle with a mass of about 2.4~PeV and with a lifetime as given in equation~(\ref{tau}) can explain the IceCube PeV neutrino events.  Note that as a result of neutrino oscillations, all neutrino flavors will contribute equally to the final signal, independent of the original flavor structure of the dark matter decays.

\section{Effective Operators}
\label{sec: effective operators}

Here we list all operators which might lead to a high energy monochromatic neutrino line from dark matter decays. In table \ref{BigTable}, we show possible dark matter candidates, defined by standard model SU(2)$_L$ and U(1)$_Y$ quantum numbers. We only list candidates which have a leading decay operator to two standard model particles, including at least one neutrino. We exclude cases in which there is an alternate decay mode through an operator of lower dimension,\footnote{In cases in which the dark matter particle carries hypercharge (case 1, 4, and 5 in table \ref{BigTable}), a Dirac mass partner is required. We only include in the table operators of lowest dimension when considering all operators allowed for either member of the Dirac pair.} or which require decays to additional non-standard model particles. Note that case 1 is a slight exception to this rule, since a decay through a lower dimension operator $H^\dagger \phi H^c$ is possible, but we include this case since the two Higgs decay may be forbidden by lepton number. In the final column of the table, we give the coefficient for the operator required to explain the two IceCube events based on the flux in the previous section.

\begin{table}[t]
\begin{center}
{\small
\begin{tabular}{c|ccc|c|c}
Case & Spin & SU(2)$_L$ & U(1)$_Y$ & Decay Operator
& Coefficient for IceCube Data \\
\hline
1. & 0 & 3 & 1 & $\bar{L}^c \phi L$ & 9.5 $\times$ 10$^{-30}$ \qquad~~~~~ \\
2. & 1/2 & 0 & 0 & $\bar{L} H^c \psi$ & 2.7 $\times$ 10$^{-29}$ \qquad~~~~~ \\
3. & 1/2 & 3 & 0 & $\bar{L} \psi^a \tau^a H^c$ & 3.8 $\times$ 10$^{-29}$ \qquad~~~~~ \\
4. & 1/2 & 2 & $-1/2$ & $\bar{L} F \psi$ & 5.6 $\times$ 10$^{-30}$ (PeV$^{-1}$) \\
5. & 1/2 & 3 & $-1$ & $\bar{L} \psi^a \tau^a H$ & 2.7 $\times$ 10$^{-29}$ \qquad~~~~~ \\
6. & 1 & 0 & 0 & $\bar{L} \slashed{V} L$ & 3.3 $\times$ 10$^{-29}$ \qquad~~~~~ \\
7. & 3/2 & 0 & 0 & $(\bar{L} iD_\mu H^c) \gamma^\nu \gamma^\mu \psi_\nu$ & 1.9 $\times$ 10$^{-29}$ (PeV$^{-1}$)
\\
\hline
\end{tabular}
}
\caption{\sl \small
Dark matter candidates and the decay operators that may lead to a mono-chromatic neutrino line signature. Here, $L$ and $H$ represent the SM lepton and Higgs doublets, respectively, while the dark matter particle is labeled by $\phi$, $\psi$, $V^\mu$ or $\psi^\mu$, depending on whether it has spin 0, 1/2, 1, or 3/2. The notation $F$ in case 4 denotes either $B_{\mu\nu} \sigma^{\mu\nu}$, $\tilde{B}_{\mu\nu} \sigma^{\mu\nu}$, $W^a_{\mu\nu} \tau^a \sigma^{\mu\nu}$ or $\tilde{W}^a_{\mu\nu} \tau^a \sigma^{\mu\nu}$ with $B_{\mu \nu}$ and $W^a_{\mu\nu}$ being the field strength tensors of the SM U(1)$_Y$ and SU(2)$_L$ gauge fields. In the final column, we give the coefficient for the operator required in order to explain the two anomalous neutrino events reported by the IceCube collaboration, assuming a dark matter particle mass of $2.4 \times 10^6$ GeV.}
\label{BigTable}
\end{center}
\end{table}

Cases in which the dark matter particle carries electric or color charge have not been included in the table. Electrically charged dark matter is severely constrained by several observations and experiments, and is required to be heavier than about $10^{12}$~GeV \cite{Gould:1989gw, Kohri:2009mi, Kamada:2013sh}, primarily by difficulties with structure formation.
Colored dark matter, similarly, must be heaver than about $10^{16}$~GeV \cite{Kawasaki:1991eu, Mack:2007xj}, with
the primary constraint coming from the possibility of overheating the Earth's core.
We have, on the other hand, included cases with non-zero hypercharge, which naively have excluded tree level $Z$-boson exchange signatures at dark matter direct detection experiments.
These constraints can be avoided, however, if there is a higher-dimension operator which induces a splitting among the components of the dark matter field in such a way that the lightest state becomes a Majorana particle.  Such a splitting is then required to be larger than the recoil energies produced at direct detection experiments. This is  in fact what occurs for the case of higgsino dark matter in supersymmetric models-- the mixing with Majorana gauginos causes the splitting.

As discussed in the introduction, all decay operators we consider in table \ref{BigTable} contain only three fields. This was done in order to ensure that a monochromatic neutrino line signal dominates over other decay modes. Operators requiring  extra insertions of Higgs vacuum expectation values to yield a monochromatic neutrino decay are not allowed, since multi-body decays with extra Higgs particles would give overly large alternate  cosmic ray signatures.

\section{Models}
\label{sec: models}

\subsection{Gravitino Dark Matter with R-Parity Violation}
\label{gravitino}

Our first example model comes from the operator listed as case 7 in table \ref{BigTable}. This operator requires the dark matter particle to have spin 3/2-- namely, to be a gravitino. Here we will show that, in an R-parity violating context, it is straightforward to obtain a monochromatic neutrino line from gravitino dark matter decays, with a mass and lifetime appropriate for explaining the PeV IceCube events.

We begin by considering the mass of the gravitino. Both the ATLAS and CMS collaborations at the LHC have reported the discovery of a Higgs boson with mass around 126~GeV~\cite{ATLAS, CMS}. The mass is somewhat heavier than one could expect in the minimal supersymmetric standard model (MSSM), but large radiative corrections to the Higgs quartic coupling~\cite{Okada:1990vk}-\cite{Arvanitaki:2012ps} can lead to a heavier lightest Higgs mass in the MSSM\footnote{It is also possible that large supersymmetry breaking terms cause some squarks to form Higgs-like bound states, hence relaxing the MSSM limits on the mass of the lightest Higgs boson~\cite{Cornwall:2012ea}.}. When the left-right mixing of scalar top quarks is negligible and $\tan \beta \simeq 2$, the typical scale of sparticle masses must be ${\cal O}(1)$~PeV assuming  simple gravity mediated SUSY breaking. Based on this observation, several concrete models have been proposed~\cite{Ibe:2011aa}-\cite{split-susy}, which are attractive from the viewpoint of the SUSY-Flavor and CP problems because all dangerous flavor changing processes are suppressed by heavy sfermion masses. Gravitino dark matter with a PeV mass is, therefore, quite consistent with the observed Higgs mass under the assumption that the gravitino is the lightest supersymmetric particle (LSP).

Let us now consider the lifetime of the gravitino, whose decay must be induced by some form of R-parity violation. Here we will consider a simple set of assumptions which will imply that the leading R-parity violating operator in the superpotential will be of the form $L H_u$, with a coefficient of order $m_{3/2}^2/M_{\rm pl}$, where $L$ is a lepton doublet of arbitrary flavor, $m_{3/2}$ is the gravitino mass, and $M_{\rm pl}$ is the Planck scale. There are several assumptions required: The first is that the R-charges of all MSSM matter fields are equal to 1, while those of the Higgses are equal to 0. Next we suppose that it is actually a $Z_3$ subgroup of $U(1)_R$ which is a symmetry of the theory, and not the full continuous $U(1)_R$. Since the gravitino mass is a spurion for R-symmetry breaking with R-charge 2, and $Z_3$ R-symmetry requires that superpotential terms have R-charge equal to 2 mod 3, we find that the R-parity violating operator $L H_u$ appears with a coefficient of $m_{3/2}^2/M_{\rm pl}$ as promised \cite{ShiraiR}. Note that other R-parity violating operators, $U D D$, $L L E$ and $Q L D$ all appear at order $m_{3/2}/M_{\rm pl}$, and therefore also with one suppression by the Planck scale. These will lead to continuum neutrino decay spectra in addition to the monochromatic line (plus continuum) obtained from the R-parity violating operator $L H_u$. Note, however that the decay rates from these other R-parity violating operators will have additional phase space suppressions due to extra final state particles, and are thus naively expected to be sub-dominant.

As a result of the $L H_u$ operator, the lifetime of the gravitino to decay into a neutrino plus a Higgs or a neutrino plus a Z boson is estimated to be~\cite{Ishiwata:2008cu}
\begin{eqnarray}
\tau_{3/2}
\simeq 192 \pi \, (M_{\rm pl}/m_{3/2})^4 \, m_{3/2}^{-1}
\simeq 10^{20} \, {\rm s},
\label{toofast}
\end{eqnarray}
where for illustration the sneutrino mass is assumed to be the same as the gravitino mass, namely 2.4~PeV, and  $M_{\rm pl} \simeq 2.4 \times 10^{18}$~GeV. This lifetime of order $10^{20}$~s is too short to be consistent with the two IceCube PeV neutrino events. 

We now note, however, the discussion leading to equation~(\ref{toofast}) potentially misses an important point. Indeed, the operator $LH_u$ carries B$-$L charge $-1$. In any case, B$-$L symmetry must be broken to allow for Majorana masses for right handed neutrinos $N_R$~\cite{seesaw1}-\cite{seesaw3}. Therefore, as is standard we may introduce $B$ and $\bar{B}$ fields carrying B$-$L charges $+1$ and $-1$, respectively in order to break this symmetry. Now one can write ${\cal W} \supset N_R N_R \langle \bar{B} \rangle^2/M_{\rm pl}$, which provides the Majorana mass term. If one assumes $M_N \sim 10^{10}$~GeV, the expectation value of the $\bar{B}$ field is $\langle \bar{B} \rangle \sim 10^{-4} M_{\rm pl}$. Then the R-parity violating operator becomes of order\footnote{Note that $B$ and $\bar{B}$ have R charge 0 according to our charge assignments, as required.}
\begin{eqnarray}
{\cal W} = (m_{3/2}^2 \langle \bar{B} \rangle /M_{\rm pl}^2) \, L H_u.
\end{eqnarray}
Because of this modification, the lifetime of the gravitino (decaying into a neutrino and either a Higgs boson or Z boson) is now about $10^8$ times longer than the lifetime in equation~(\ref{toofast}). To be more precise, the lifetime is then estimated to be
\begin{eqnarray}
\tau_{3/2}
\simeq 192 \pi \, (M_{\rm pl}/\langle \bar{B} \rangle)^2
(M_{\rm pl}/m_{3/2})^4 \, m_{3/2}^{-1}
\simeq 10^{28} \, {\rm s},
\end{eqnarray}
where the Majorana mass has been set to 10$^{10}$ GeV. This lifetime is fully consistent with the one implied by the IceCube PeV neutrino flux in equation~(\ref{tau}).

Finally, let us discuss gravitino production in the early universe and the dark matter abundance. The next-to-lightest-superpartner (NLSP) in this model will generically decay to the gravitino (plus its standard model partner) with a lifetime of order $M_{\rm pl}^2/m_{\rm NLSP}^3$. With an NLSP mass at the PeV scale this is roughly of order $10^{-5}$ seconds. The NLSP decays thus do not disrupt the successful predictions of big bang nucleosynthesis. On the other hand, the NLSP freeze-out abundance, which will then be converted into the gravitino relic abundance, will be too high by perhaps a factor of $\sim10^5$ due to the large NLSP mass. If the reheating temperature is above the NLSP mass, entropy production by a factor of $\sim10^5$ will thus be required to dilute the dark matter abundance. If the reheating temperature is below the NLSP mass on the other hand, then it is possible to produce an appropriate gravitino abundance through a small branching fraction of the inflaton into the gravitino.

\subsection{Hidden Sector Gauge Boson}
\label{hidden}

Another interesting possibility for a dark matter particle which could give a neutrino line signature at IceCube comes from case
6 in table \ref{BigTable}.  Here we require a new gauge boson $V^\mu$ with a very small coupling $\sim 10^{-28}$ to at least one standard model lepton.  What is very interesting about this case is that a coupling of this size may be obtained in a very simple and natural way.\footnote{For another
model which may be used to give a similar resulting decay operator please see \cite{BLmix}.}  In particular, let us suppose that the visible sector is part of a standard grand unified theory,
with a unification scale of $M_{\rm GUT} \sim 10^{16}$GeV.  We may take a minimal ${\rm SU}(5)$ theory with GUT symmetry broken by the vacuum expectation value of an adjoint scalar field $\Sigma$ for illustration.

Now, we consider the possibility that there is a completely hidden sector with
a new non-abelian gauge symmetry, broken at the PeV scale.  Let us take this gauge symmetry to be ${\rm SU}(2)$ for simplicity, and suppose that it is completely broken by the vacuum expectation value (vev) of a scalar field $\Phi$ in the fundamental representation.  Because both the visible and hidden sector gauge symmetries are fundamentally non-abelian, dimension four kinetic mixing between their respective field strengths coming from an operator $\sim F_{\mu\nu}V^{\mu\nu}$ is forbidden, where now $F^{\mu\nu}$ and $V^{\mu\nu}$ are taken to be the $SU(5)$ and hidden sector field strengths, respectively.
However, after gauge symmetry breaking, such mixing is induced  by Planck suppressed operators, even if there are 
no new particles carrying both visible and hidden quantum numbers.
 In the present example, the minimal Planck suppressed operator which results in mixing between the visible and hidden gauge bosons is given
 by\footnote{Operators with different combinations of  $\Phi$ and $\Phi^\dagger$ are  similarly allowed.}
\begin{equation}
\mathcal{L}\supset \frac{1}{M_{\rm pl}^3}\Sigma F_{\mu\nu} \Phi^\dagger   V^{\mu\nu} \Phi.
\label{mixing}
\end{equation}
This leads to a kinetic mixing between hypercharge and the lightest new gauge boson of order $\frac{\vev{\Phi}^2 \vev{\Sigma}}{M_{\rm pl}^3} \sim \frac{{\rm PeV}^2 M_{\rm GUT}}{M_{\rm pl}^3} \sim 10^{-28}$.  The hidden gauge boson will then obtain a coupling to the standard model leptons
of the right order to explain the IceCube data!  Of course, we are assuming here that there are no light hidden sector particles into which the hidden gauge boson may rapidly decay.

Note that in the absence of supersymmetry, this model introduces a new hierarchy problem for the mass of the scalar field $\Phi$.  There is also a dangerous allowed quartic coupling between $\Phi$ and the visible sector Higgs boson, which would lead to very rapid $V^\mu$ decays to Higgs bosons.  There is, however, no obstacle to implementing the model in a supersymmetric framework, and doing so can prevent the quadratic divergence of the $\Phi$ mass, as well as forbid the $\Phi/$Higgs quartic coupling.  On the other hand, there is one additional type of dangerous operator which supersymmetry cannot forbid.  This is an operator of the form
\begin{equation}
 \frac{1}{M_{\rm pl}^2} \Phi^\dagger V^\mu \Phi H^\dagger \partial_\mu H,
\label{danger}
\end{equation}
 which may be generated by a Kahler potential term
$\frac{1}{M_{\rm pl}^2} \Phi^\dagger \Phi H^\dagger H$ and which results in $V^\mu$ decays to two Higgses.   Note that the operator (\ref {mixing}) which leads to the monochromatic neutrino line is suppressed by an additional factor of $\frac{M_{\rm GUT}}{M_{\rm pl}}$ compared to (\ref{danger}).  We thus require that the new operator  be suppressed by a factor of about 100-1000 beyond the naive estimate in (\ref{danger}) of $\frac{1}{M_{\rm pl}^2}$ in order that the monochromatic neutrino line is the dominant cosmic ray signature.  Note that by gauge invariance (\ref{danger}) is necessarily accompanied by a factor of the hidden gauge coupling constant, while the operator (\ref{mixing}) may not be, so that a somewhat small gauge coupling may be able to account for some or all of the required suppression.

Similarly to the gravitino case, an appropriate dark matter relic abundance for the hidden gauge boson may be obtained through non-thermal production.  For example, we may suppose that the inflaton decays with an appropriate branching fraction into the hidden sector, while also reheating the
visible sector.

\subsection{A singlet fermion in an extra dimension}
\label{smallyukawa}

Here we point out that in the context of an extra dimension, it is straightforward to obtain a highly suppressed coupling such as that needed for a monochromatic neutrino line from dark matter decay.  In particular, we may consider a scenario to produce the operator of case 2 in table \ref{BigTable}.

Suppose that there exists an $S_1/Z_2$ orbifolded fifth dimension separating two branes.  One of these branes, at $y=0$, hosts all of the
standard model fields, while the other, at $y=\ell$, hosts a Majorana mass term for the right handed part of a singlet Dirac fermion $\Psi$ which  propagates in the bulk.
In addition, $\Psi$ has a  mass term in the bulk and a Yukawa coupling to a lepton and the Higgs on the standard model brane.  
The zero mode of the right handed part of this bulk fermion may then be exponentially suppressed on the standard model brane, taking the form\footnote{The zero mode for the left handed part of $\Psi$ is set to zero by choosing it to be odd under the $Z_2$ orbifold symmetry as usual.}
\begin{eqnarray}
\Psi_R^{(0)}(y, x) =
\sqrt{\frac{2m}{e^{2m\ell} - 1}}
\frac{1}{\sqrt{M_*}} e^{m y} \psi_R^{(4D)}(x) \equiv \varepsilon e^{m y} \psi(x),  
\label{relation}
\end{eqnarray}
where $M_*$ is the fundamental scale related to the four-dimensional Planck scale by $M_{\rm pl}^2 \;=\; M_*^3 \ell.$ Here we have written the action for the zero mode of $\Psi$ as
\bea
S&=& \int d^4x \,dy  \left\{ M_*\left(i\bar{\Psi}^{(0)} \Gamma^A \partial_A \Psi^{(0)} + m \bar{\Psi}^{(0)} \Psi^{(0)} \right)
\right.\non\\ 
&& \left.
+ \left[ \delta(\ell - y) M_R  \bar{\Psi}^{(0)c}_{R} \Psi^{(0)}_{R}
 + \delta(y) \lambda \bar{\Psi}^{(0)}_{R}  L \, H + {\rm h.c.}  \right] \right\}.
\label{action}
\eea
It is then straightforward to choose $M_R$ to be 2.4~PeV to explain the energies of the IceCube events, while due to exponential suppression
$\varepsilon$ may be taken to be of order $10^{-29}$ to yield an appropriate $\psi$ dark matter lifetime even if $\lambda$
is of order 1.

As in the previous examples we have discussed, inflaton decays may yield an appropriate dark matter relic abundance.
In this case there is also another interesting possibility- namely, we may take the dark matter particle to carry gauged $B-L$ symmetry (along with the standard model fermions, and two more right handed neutrino-like states for anomaly cancellation), so that the mass $M_R$ is only produced after spontaneous $B-L$ breaking.  In this case, $B-L$ interactions in the
early universe may be used to produce the needed $\psi$ relic density.
The correct abundance of dark matter can be attained if the reheat temperature $T_R$ 
is below the scale at which the U(1)$_{B-L}$ gauge symmetry is restored, and also below the temperature at which $\psi$ particles would come into thermal equilibrium through gauge interactions.  
The population of $\psi$ particles can be produced in processes $ll \rightarrow \psi \psi$ mediated by the heavy U(1)$_{B-L}$ gauge boson.  The resulting density to entropy ratio can 
be estimated as in Ref.~\cite{Kusenko:2010ik}: 
\bea
Y_{\psi}\; \equiv\; \frac{n_{\psi}}{s}  &\sim& \left.\frac{\la \sigma v \ra n_f^2/ \tilde {H}}{\frac{2 \pi^2}{45}g_* T^3}\right|_{T=T_R} \non\\
			&\sim & 10^{-16} \lrfp{g_*}{10^2}{\frac{3}{2}} \lrfp{M_{B-L}}{\GEV{18}}{-4} \lrfp{T_R}{5 \times \GEV{13}}{3},
\eea
where $\tilde H$ is the Hubble parameter, $g_*$ is the number of relativistic degrees of freedom at the time of reheating, $\la \sigma v \ra \sim T^2/M_{B-L}^4$ is the production cross section,
$n_f \sim T^3$ is the number density of standard model fermions in the plasma, and the first equality is
evaluated at  reheating.  Numerical solution of the Boltzmann
equation gives a value consistent with this result~\cite{Khalil:2008kp}.  The dark
matter mass density is then
\begin{eqnarray}
\Omega_{\rm dark}=0.2 \times \left(\frac{m_{\psi}}{2.4\times 10^6\, {\rm GeV}}\right) \left( \frac{Y_{\psi}}{1.5 \times 10^{-16}} \right).
\label{Omegadm}
\end{eqnarray}

\section{Discussion and Future Prospects}
\label{sec: future}

In this paper we have catalogued all of the operators which may lead to decays of PeV dark matter particles into monochromatic neutrino lines, and have given examples of models which may lead to appropriate decay rates to explain the two anomalous events recently reported by the IceCube collaboration. Here we would like to highlight an interesting feature of our analysis: For all of the operators that we have discussed in this paper one actually obtains also a lower energy continuum of cosmic ray neutrinos in addition to the monochromatic neutrino line. These are produced since in every case there are necessarily alternate primary decay products-- in addition to the primary neutrinos-- which include Higgses, W-bosons, Z-bosons and charged leptons, whose decays in turn produce neutrinos at lower energies.
For example, in the hidden gauge boson model discussed in section \ref{hidden}, the vector dark matter particle decays into all standard model particles carrying hypercharge. In particular, we will obtain decays to muons and tau leptons leading to a continuum neutrino signature. In the gravitino model of section \ref{gravitino} and the singlet fermion model in section \ref{smallyukawa}, there are necessarily decays to W-boson + charged lepton which produce continuum neutrinos, in addition to those produced from the Higgs and Z-boson final state particles in the primary neutrino decays. For all cases we have considered in this paper, these final states leading to continuum neutrino signals have a similar branching fraction to the monochromatic neutrino events which have been our primary interest. We therefore have the important result that if the IceCube PeV events are due to dark matter decays, then there should also be a continuum of excess lower energy events that can also be discovered in the sub-PeV region.\footnote{Note that for the dominantly monochromatic neutrino spectra which 
we are considering in this paper, one necessarily also obtains  a continuum of soft
neutrinos via electroweak bremsstrahlung, independent of any model building considerations.  However, such bremsstrahlung induced neutrinos have a spectrum which is too soft to be observable at IceCube.  In particular, they only contribute to the continuum spectrum in an appreciable way at low energies where they are dwarfed by the atmospheric background.  The decays of primary decay products thus give the most important contribution to the neutrino continuum.}

\begin{figure}[t]
\begin{center}
\includegraphics[width=0.7\hsize]{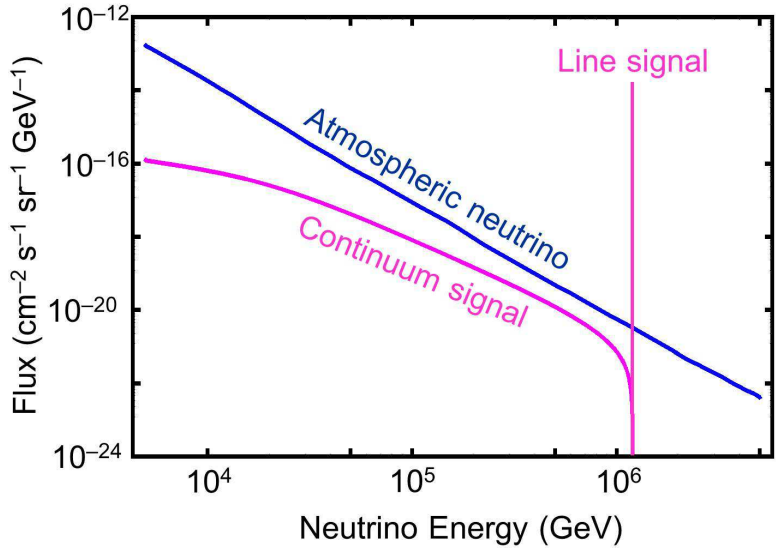}
\caption{\sl Line and continuum neutrino signals from PeV dark matter decays.}
\label{fig: neutrino signals}
\end{center}
\end{figure}

While the precise size and shape of this continuum is model-dependent, qualitatively it always has a similar form. In figure \ref{fig: neutrino signals} we show both line and continuum signals assuming that the partial decay width of the continuum signal is twice that of the line signal. This corresponds to the cases of either the gravitino model or the singlet fermion
model discussed in the text.  The combined atmospheric neutrino background (including those from prompt decays)~\cite{icecube1} is also shown for comparison. The continuum flux was calculated using the method adopted in reference~\cite{Birkel:1998nx}, and is based on the contribution from hadronic cascade decays of SM particles. In addition, we can also expect another contribution from  leptonic decays, but this is not included in the figure for simplicity.
Note that in both the gravitino and singlet fermion cases we also have direct decays into a W-boson plus a charged lepton $l$, with the flavor of the lepton being model dependent.
Error introduced by our approximation of dropping leptonic decays will be negligible for the cases of $l = e$ or $l = \tau$, 
while if $l = \mu$, the continuum spectrum in the sub-PeV region will be somewhat enhanced.

Let us note one special case in which the prediction of appreciable
 continuum neutrinos may be avoided-- namely, one may consider the possibility that dark matter decays produce neutrinos along with a new hidden sector particle, rather than additional standard model ones.  For an interesting example, we may take a scalar dark matter particle $\phi$ which decays into two hidden sector singlet fermions $\psi$. If $\psi$ actually mixes with standard model neutrinos, then this will lead to decays of $\phi$ to $\psi$ plus a neutrino, without an appreciable continuum neutrino signal. We will give details of a split seesaw model which realizes this scenario in appendix A.

One might wonder in addition about the possibility of other types of cosmic ray signatures from decay products in our models, such as gamma rays or antiprotons. Unfortunately these are unlikely to be detectable in the foreseeable future. The reason is the following: First, backgrounds of diffuse gamma rays and cosmic ray antiprotons have fluxes whose energy spectra are softer than $1/E^2$, due to their production by cosmic-ray protons. On the other hand, gamma rays and antiprotons from dark matter decays have fluxes whose energy spectra are harder than $1/E^2$ (typically going as $1/E$). This is because the signal spectra are essentially determined by the fragmentation functions of dark matter decays and these must be harder than $1/E^2$, otherwise their integrals over energy will diverge. As a result, the ratio of the signal flux to the background flux becomes smaller at smaller energies. Moreover, both gamma rays and antiprotons are now observed at most up to 1 TeV in energy, making detection difficult. This situation may be clearly seen in reference~\cite{Murase:2012xs} for the gamma-ray case, where it was shown that near future gamma-ray observations can cover dark matter lifetimes at most up to $10^{27}$ seconds.

\vspace{.75cm}
Note Added:  After this work was completed a paper by the IceCube collaboration discussing these events was released \cite{Aartsen:2013bka}. The event energies were adjusted slightly compared to those used here, but there is no significant impact on our results.
\vspace{.75cm}

\noindent
{\bf Acknowledgments}
\vspace{0.1cm}\\
\noindent
The authors thank F. Halzen and A. Ishihara for very helpful discussions about the IceCube events. A.K. was supported by DOE Grant DE-FG03-91ER40662. S.M. and T.T.Y. were supported by the Grant-in-Aid for Scientific research from the Ministry of Education, Science, Sports, and Culture (MEXT), Japan (No. 23740169 for S.M. and No. 22244021 for S.M. \& T.T.Y.). This work was also supported by the World Premier International Research Center Initiative (WPI Initiative), MEXT, Japan.

\appendix
\section{A Split Seesaw Model}

Here we will discuss a model which is outside of the main line of argument in the text for two reasons:  the first is that
there is an additional hidden  sector particle  in the decay final state, and the second is that the decay may not 
be thought of  as due to a single effective operator, since it is a result of a mixing between two low mass particles. 
As mentioned in the discussion section, the basic idea is to have a scalar dark matter particle $\phi$ decaying to two light
hidden sector fermions $\psi$ through a (highly suppressed) $\phi \psi \psi$ interaction, and also require that $\psi$ has some
mixing with standard model neutrinos.  We will now show that such a situation may be obtained in a split seesaw framework
in an extra dimension \cite{Kusenko:2010ik}, in which the fermion $\psi$ can literally be a right handed neutrino in the sense that it
leads to a seesaw neutrino mass in the standard model \cite{seesaw1,seesaw2,seesaw3}, even though $\psi$ itself will be very light.\footnote{We of course need more than one non-zero neutrino mass in the standard model sector, and thus require more than one right handed neutrino.  This will not concern us here as 
a single $\psi$ field is sufficient for our present purpose.}

The basic setup is similar to the one used in section \ref{smallyukawa}.  We again put standard model fields on a $y=0$ brane
in an extra dimension, with a $\Psi$ field propagating in the bulk as in that section, and with a zero mode wavefunction peaked on the brane at $y=\ell$.  Again we also put a Yukawa coupling between $\Psi$ $L$ and $H$ on the standard model brane leading to an interaction  $\varepsilon \lambda \psi L H$, where we are continuing to use the notation of section \ref{smallyukawa}.
 A difference here however, is that  we will now put the Majorana mass $M_R$ for $\Psi$ on the standard model brane rather than the  $y=\ell$ brane.  As a result, $\psi$ will obtain a highly suppressed mass of $\varepsilon^2 M_R$. 
An interesting result-- and the original motivation for the split seesaw framework-- is that a seesaw mass  is then obtained 
for a standard model neutrino which is interestingly independent of the wavefunction suppression factor $\varepsilon$, with
$m_\nu = \lambda^2 v^2/M_R$.  

Finally, we introduce a new scalar field $\phi$ living on the standard model brane, with a Yukawa coupling to $\Psi$ resulting in an interaction of size $g \varepsilon^2 \phi \psi \psi$.  $\phi$ will be our dark matter particle, and thus we choose its mass to be
$2.4$ PeV.

After electroweak symmetry breaking, one obtains a potentially large mixing between $\psi$ and a neutrino $\nu$ of $\frac{\lambda v}{\varepsilon M_R}$, where $v$ is the standard model Higgs vev.  While the primary decay mode of $\phi$ will be to two $\psi$ particles, as a result of the mixing, $\phi$ may also decay to $\psi \nu$  with a lifetime of order $\left(\frac{10^{-28}}{g \varepsilon^2}\right)^2 \times 10^{28} {\rm s}$, where we have taken the mixing angle to be of order 1.   
Obtaining an appropriate neutrino mass $m_\nu$ with $\lambda$ also of order 1 requires $M_R$ to be of order $10^{16}$~GeV as usual.
Finally let us point out that, as was discussed in section \ref{smallyukawa}, an interesting possibility for producing the dark matter abundance results if one assumes that $\phi$ carries gauged $U(1)_{B-L}$ charge, so that high temperature $B-L$ interactions produce the relic $\phi$ particles. The estimate for the resulting relic density is analogous to that in section \ref{smallyukawa}.\footnote{In the original split-seesaw papers $\psi$ was a dark matter candidate with a keV mass  which could explain pulsar kicks ~\cite{Kusenko:1996sr, Kusenko:2009up}.  However, the keV mass scale was not a definitive prediction of the model, which essentially creates a ``democracy of scales'': Majorana masses of different orders of magnitude can arise from the exponential suppression employed in this model.  For example, two degenerate right-handed neutrinos with GeV masses as used in the $\nu$MSM~\cite{Asaka:2005an} can be accommodated in this scenario as well.} 


\end{document}